\documentclass[onecolumn,superscriptaddress]{revtex4}
\usepackage{graphicx}% Include figure files
\usepackage{dcolumn}% Align table columns on decimal points
\usepackage{bm}% bold mats
\usepackage{amsmath}
\usepackage{epsfig}
\begin{document}

\title{Search for Jacobi shape transition in A $\sim30$ nuclei}

\author{Balaram Dey}
\affiliation{Department of Nuclear and Atomic Physics, Tata Institute of Fundamental Research, Mumbai-400005, India}
\author{C. Ghosh}
\affiliation{Department of Nuclear and Atomic Physics, Tata Institute of Fundamental Research, Mumbai-400005, India}
\author{Deepak Pandit}
\affiliation{Variable Energy Cyclotron Centre, 1/AF-Bidhannagar, Kolkata-700064, India}
\author{A.K. Rhine Kumar}
\affiliation{Department of Physics, Cochin University of Science and Technology, Cochin-682022, Kerala, India.}
\author{S. Pal}
\affiliation{Pelletron Linac Facility, Tata Institute of Fundamental Research, Mumbai-400005, India}
\author{V. Nanal}
\email[e-mail:]{nanal@tifr.res.in}
\affiliation{Department of Nuclear and Atomic Physics, Tata Institute of Fundamental Research, Mumbai-400005, India}
\author{R.G. Pillay}
\affiliation{Department of Nuclear and Atomic Physics, Tata Institute of Fundamental Research, Mumbai-400005, India}
\author{P. Arumugam}
\affiliation{Department of Physics, Indian Institute of Technology, Roorkee-247667, India}
\author{S. De}
\affiliation{Nuclear Physics Division, Bhabha Atomic Research Centre, Mumbai-400085, India}
\author{G. Gupta}
\affiliation{Department of Nuclear and Atomic Physics, Tata Institute of Fundamental Research, Mumbai-400005, India}
\author{H. Krishnamoorthy}
\affiliation{Indian Neutrino Observatory, Tata Institute of Fundamental Research and Homi Bhabha National Institute, Mumbai-400085, India}
\author{E.T. Mirgule}
\affiliation{Nuclear Physics Division, Bhabha Atomic Research Centre, Mumbai-400085, India}
\author{Surajit Pal}
\affiliation{Variable Energy Cyclotron Centre, 1/AF-Bidhannagar, Kolkata-700064, India}
\author{P.C. Rout}
\affiliation{Nuclear Physics Division, Bhabha Atomic Research Centre, Mumbai-400085, India}


\date{\today}

\begin{abstract}
This paper reports the first observation of the Jacobi shape transition in $^{31}$P  using high energy $\gamma$-rays from the decay of giant dipole resonance (GDR) as a probe. The measured GDR spectrum in  the decay of $^{31}$P  shows a distinct low energy component around 10 MeV, which is a clear signature of Corioli's splitting in a highly deformed rotating nucleus. 
 Interestingly,   a self-conjugate $\alpha$-cluster nucleus $^{28}$Si, populated at similar initial excitation energy and angular momentum, exhibits a vastly different GDR line shape. Even though the angular momentum  of the compound nucleus $^{28}$Si is higher than the  critical angular momentum  required for the Jacobi shape transition, the GDR lineshape is akin to a prolate deformed nucleus. Considering the present results for $^{28}$Si and similar observation recently reported in $^{32}$S, it is proposed that the nuclear orbiting phenomenon exhibited by $\alpha$-cluster  nuclei hinders the Jacobi shape transition.  The present experimental results suggest a possibility to investigate the nuclear orbiting phenomenon using high energy $\gamma$-rays as a probe.
\end{abstract}
%\keywords{}
\maketitle
Many body quantum systems like atomic nuclei provide a unique opportunity to explore a variety of phenomena arising due to interplay of different physical processes, particularly at high excitation energy ($E^*$) and angular momentum ($J$). One such interesting phenomenon is the Jacobi shape transition, where beyond a critical angular momentum ($J_C$),  an abrupt shape change from non-collective oblate shape to collective triaxial or prolate shape takes place~\cite{beri}. 
The study of exotic Jacobi shapes  in nuclei has been a topic of considerable interest~\cite{myers, dudek}.  The Jacobi shape transition is expected to occur
in light and medium mass nuclei, where high rotational frequencies are achieved before the excited nucleus can undergo fission.    Further, it is expected that the Jacobi shape transition should be a common feature over a wide range of nuclei.
Experimentally, the Jacobi shape transition has been  observed in a few light mass nuclei $A \sim$ 45 \cite{kici,maj,drc1,dipu1} via the $\gamma$-decay of giant dipole resonance (GDR).  It is known that the GDR is the cleanest, and hence most extensively used, probe to study the properties of nuclei at high temperature ($T$) and $J$ \cite{DRCrev}. The GDR can be understood macroscopically as an out-of-phase oscillation between protons and neutrons, and microscopically in terms of coherent particle-hole excitations. The GDR $\gamma$-emission occurs  at the early stage of compound nucleus (CN) decay and can probe the nuclear shape. The GDR components corresponding to vibration along and perpendicular to the axis of rotation are differently affected by the Corioli's force.  As a result the GDR strength function splits into multiple components with a  narrow well separated peak around 8-10 MeV~\cite{kici}, which is an unambiguous  signature of  the Jacobi shape transition.
It should be mentioned that the search for  Jacobi shapes has also been made through studies of quasi-continuum gamma radiation ~\cite{ward}. However,  indications of highly deformed shapes could not be uniquely ascribed to the Jacobi shape. 

While many of the observed features of nuclei at high $E^*$, $J$ can be understood in terms of rotating liquid drop model (RLDM) and mean field approach, it is well known that nuclei also exhibit cluster structure ~\cite{freer1,von,Aradhana,ebran}. The influence of clustering in the stellar nucleosynthesis has been a long-standing question in nuclear astrophysics \cite{hoyle, rana, freer2}. 
Nuclear orbiting phenomena involving formation of a long-lived dinuclear molecular complex, with a strong memory of the entrance channel,
 has been observed in reactions involving self conjugate $\alpha$-cluster nuclei~ \cite{sand, kundu}. Such an orbiting dinuclear system can attain complicated exotic shapes as compared to a shape equilibrated compound nucleus \cite{freer1,sand,tani,ichi, beck}. 
Interestingly,   recent studies of GDR spectrum from  the $^{32}$S nucleus populated with $J > J_C$ in the reaction $^{20}$Ne+$^{12}$C, did not show evidence of the Jacobi shape transition \cite{dipu1,dipu2} and the result was interpreted in terms of the formation of $^{16}$O + $^{16}$O  molecular structure  in a superdeformed state of $^{32}$S.  
The observation of a narrow  resonance in $^{24}$Mg+ $^{24}$Mg at J = 36 $\hbar$ ~\cite{salsac} was interpreted in terms of the highly deformed shape  corresponding to a molecular state, but no clear signature of the Jacobi shape transition was observed.
 
Therefore, experimental studies of  the exotic shapes of different nuclei with  $J > J_C$  are crucial to understand the different mechanisms like the nuclear orbiting, cluster formation and Jacobi shape transition. 
The aim of the present study is to investigate the deformed shapes of a $\alpha$-cluster ($^{28}$Si) and a non-$\alpha$-cluster ($^{31}$P) nuclei at high $J$ using the GDR as a probe. This work also addresses the open question whether the Jacobi shape transition is a general phenomenon in light mass nuclei.

The experiments were performed using pulsed beams of $^{19}$F (at $E_{lab}$=127 MeV) and $^{16}$O (at $E_{lab}$=125 MeV) from the Pelletron Linac Facility (PLF),  Mumbai bombarding a self-supporting $^{12}$C target (400 $\mu$g/cm$^2$). 
The high energy $\gamma$-rays in the region of 5-30 MeV were measured using an array of seven close packed hexagonal BaF$_2$ detectors (each 20 cm long with face-to-face distance of 9 cm) mounted at 125$^0$ with respect to the beam direction and at a distance of 57 cm from the target position. A  14-element BGO multiplicity filter (hexagonal, 6.3 cm long and 5.6 cm face-to-face) was mounted in a castle geometry surrounding the target ($\sim$60$\%$ efficiency at 662 keV), for measuring the multiplicity ($M$) of low energy discrete $\gamma$-rays to extract the angular momentum information.  The BaF$_2$ array was surrounded by an annular plastic detector which was used as a cosmic ray veto. Detector arrays, upstream collimators and the beam dump (kept at $\sim$2 m from the target) were suitably shielded to minimize the background. The anode output of individual BaF$\rm _2$ detector was integrated in two different gates of width 200 ns ($E_{\rm short}$) and 2 $\mu$s ($E_{\rm long}$) for pileup rejection using pulse shape discrimination (PSD) and energy measurement, respectively. The time-of-flight (TOF) of each BaF$_2$ detector with respect to the RF pulse was used to reject neutron events. For each event $E_{\rm short}$, $E_{\rm long}$, BaF$\rm _2$-TOF of  each BaF$_2$ detector were recorded together with the fold F (number of BGO detectors fired for $E_{th}>$ 120 keV within a 50~ns coincidence window) and  BGO-TOF with respect to the RF pulse ~\cite{lamps}. 
The energy calibration of the  BaF$\rm _2$ detector array was obtained  using low energy radioactive sources and was linearly extrapolated to high energies. The gain stability of the BaF$\rm _2$ detectors was found to be  within $\pm$1$\%$. The beam induced background contributions were also monitored with a blank target frame and were found to be negligible.  Further details of the experimental setup can be found in Ref. ~\cite{ghosh17}.

\begin{figure}
\begin{center}
\includegraphics[height=5.0 cm, width=7.0 cm]{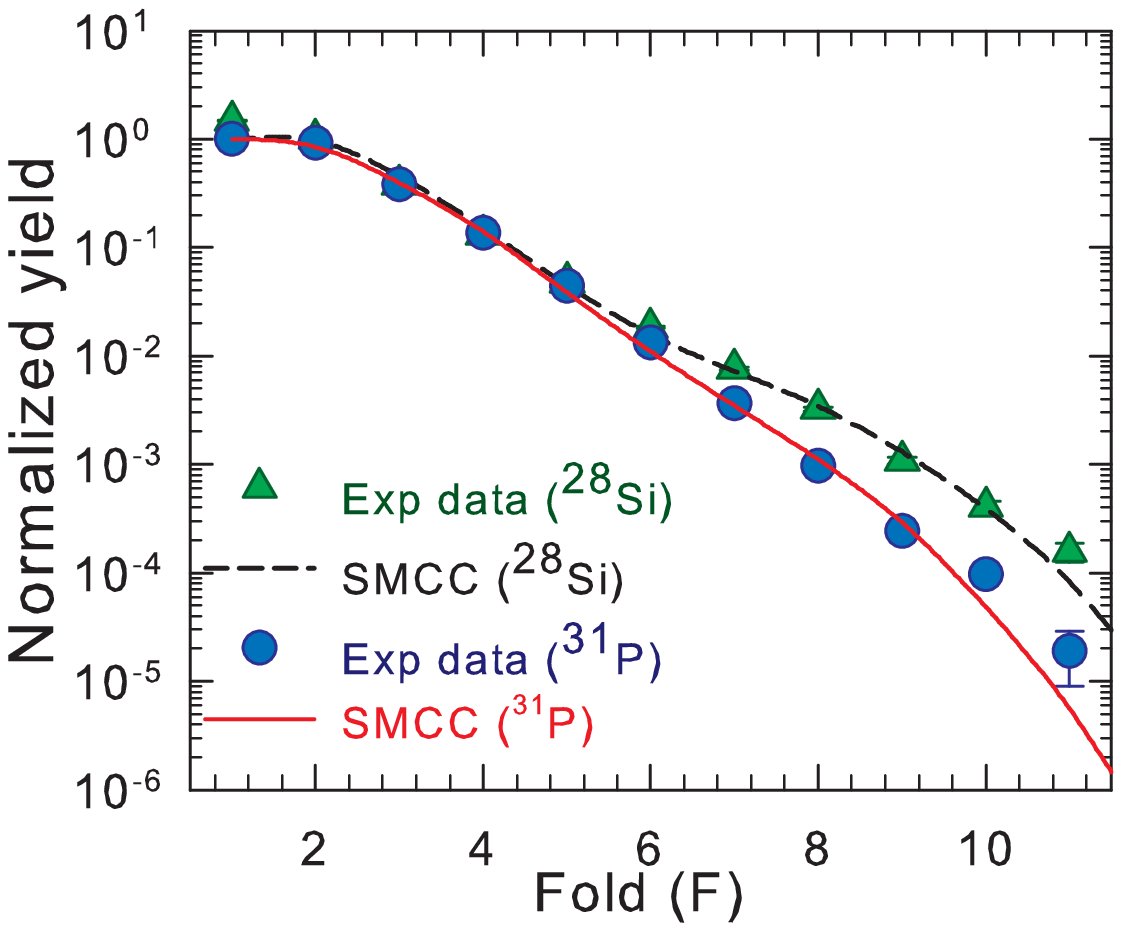}
\caption{\label{fig1} (Color online)  Experimental fold distribution (symbol) together with that from the SMCC calculations  (line) for $^{19}$F+$^{12}$C and  $^{16}$O+$^{12}$C reactions.}
\end{center}
\end{figure}
 
The high energy $\gamma$-ray spectra for different folds are generated in offline analysis after incorporating corrections due to  chance coincidence and Doppler effect arising from the finite recoil velocity of the residues. The $\gamma$-ray spectra for $F\ge4$ were analyzed within the statistical model framework using the code SMCC~\cite{chakraborty2} to extract the GDR parameters,  $\langle T \rangle$ and   $\langle J \rangle$ following the procedure in Ref.~\cite{ghosh17}.
In both systems, the data corresponding to  lower folds ($F \le 3$) were not considered as it can have contributions from radioactivity and extraneous background. 
The optical model parameters are taken from Ref. ~\cite{perey1,perey2,mcfa} and Ignatyuk level density prescription~\cite{igna} is used with $\it \tilde{a}$ = $\it A$/7 $\rm MeV^{-1}$~\cite{debasish}. The effective moment of inertia  is assumed to be $I_{eff}$ = $I_0$(1 + $\delta_1J^2$ + $\delta_2J^4$), where $I_0$(= $\frac{2}{5}A^{5/3}r_0^2$) is the rigid-body moment of inertia, $r_0$ is radius parameter, $\delta_1$ and $\delta_2$ are deformation parameters. The residue spin distribution ($J_{res}$) is calculated starting from the standard $J_{CN}$ distribution and is converted to the multiplicity $M$ using the relative decay probability ($P_r$) of dipole and quadrupole transitions as a parameter. 
The $M$ distribution is then converted to the fold ($F$) distribution incorporating the BGO array efficiency and crosstalk probability as described in Ref. ~\cite{chakraborty3}. All three parameters, namely, $P_r$, $\delta_1$ and $\delta_2$ are varied to fit the experimentally observed fold distribution. The fold distributions thus calculated with the SMCC for both systems are  shown in Fig.~\ref{fig1} together with the data. 
It is important to note that both reactions are studied in the same setup and with a similar analysis, to rule out any systematic factors that could affect the data.

\begin{figure}
\begin{center}
\includegraphics[height=8.5 cm, width=7.5 cm]{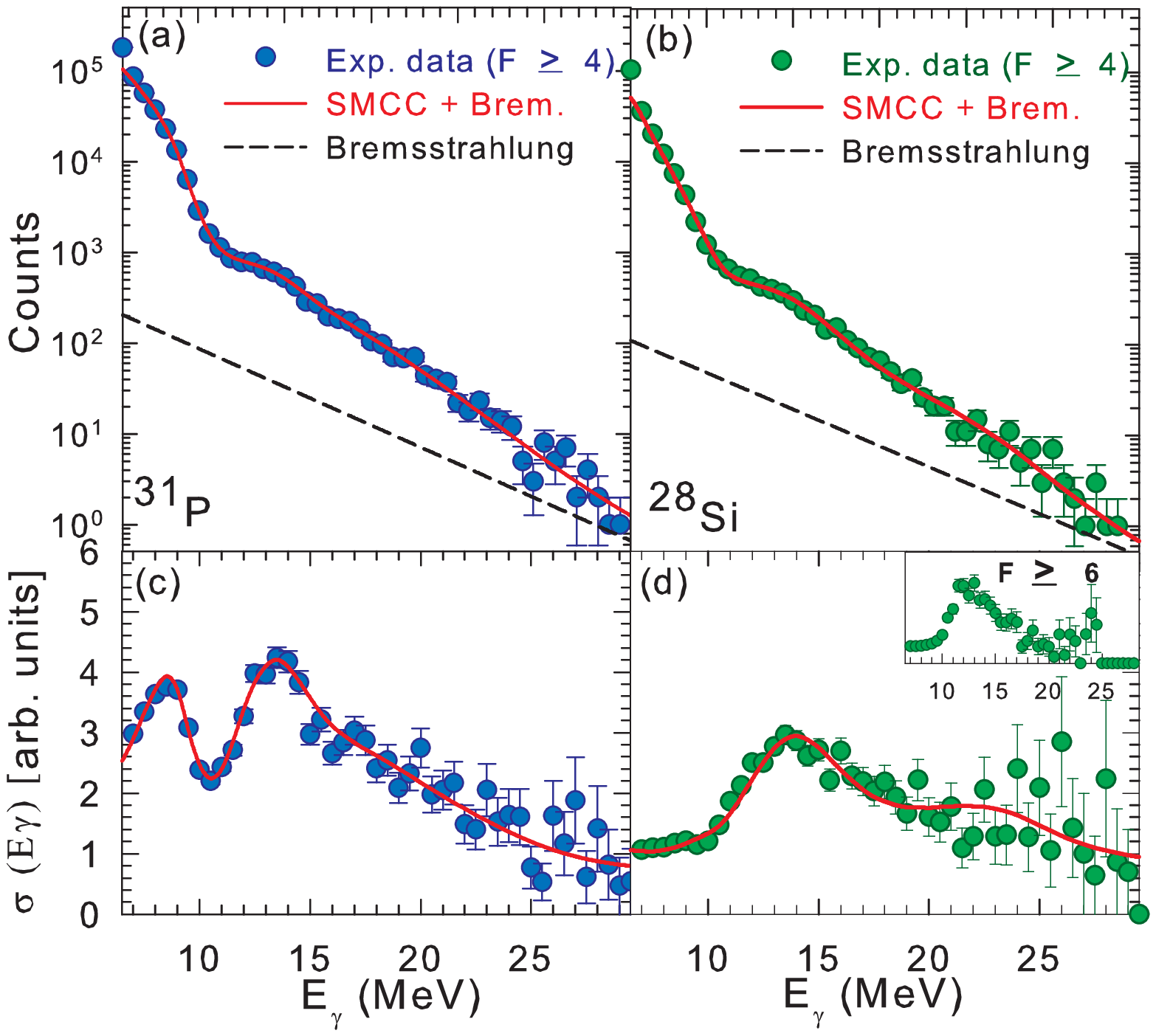}
\caption{\label{fig2} (Color online) Fold gated high energy $\gamma$-ray spectra (symbols) with the best fit SMCC calculation (line) for (a) $^{31}$P  and (b) $^{28}$Si; corresponding divided plots are shown in panel (c) and (d). Divided plot for $^{28}$Si with $F \ge 6$ is shown as an inset in panel (d).}
\end{center}
\end{figure} 

The $\gamma$-ray spectrum in the SMCC is calculated assuming $\sim$100$\%$ TRK sum rule and the GDR line shape is taken as a sum of multiple (2 to 5) Lorentzian components.
A bremsstrahlung contribution is computed  from systematics~\cite{nifnecker} as (e$^{-E_{\gamma}/E_0}$) with $E_0$ = 1.1[($E_{lab}$- $V_{c}$)/$A_{p}$]$^{0.72}$, where $E_{lab}$, $V_{c}$ and $A_{p}$ are the beam energy, Coulomb barrier and the projectile mass, respectively. The bremsstrahlung spectrum folded with the detector response function was added to the calculated GDR spectrum for comparison with data.
 The goodness of the fit is achieved by $\chi ^2$ minimization and visual inspection in the energy range of $E_{\gamma}$ = 7 - 25 MeV. Fold gated  high energy $\gamma$-ray spectra and the divided plots (generated using a $\gamma$-ray spectrum calculated with an
arbitrary constant dipole strength of 0.2 W.u.  folded with the BaF$_2$ array response)
together with the best fit statistical model calculations for both  $^{31}$P and $^{28}$Si are shown in Fig.~\ref{fig2}. 
The GDR spectrum of $^{31}$P could be not be fitted with prolate/oblate shape (2-component Lorentzian function) or a triaxial shape (3-component Lorentzian function). The observed spectrum has five Lorentzian components, resulting from the Jacobi shape transition. To restrict the fitting window (for large no. of parameters), initial values for centroid energy ($E_i$), width ($\Gamma_i$) and strength ($S_i$) were  taken from Ref.~\cite{cori-fit} and then varied individually within a limited range to achieve the best fit. In case of  $^{28}$Si, a two component strength function corresponding to a prolate shape describes the data well.
\begin{table}[!ht]
\caption{\label{table1} Best fit GDR parameters from the SMCC analysis.}
\vspace{5mm}
%\begin{ruledtabular}
\begin{center}
\begin{tabular}{|c|c|c|c|c|c|}
\hline 
System & $\langle J \rangle (\hbar) $ & $\langle T \rangle $ (MeV)&
 $E_{GDR} (\rm MeV)$  &$\it \Gamma_{GDR} (\rm MeV)$ & $S_{GDR}$\\ \hline
$^{31}$P& 22(6)   & 2.2(3) & 9.1(1)  & 2.2(1) & 0.18(2)\\ 
        &         &        & 14.2(3)  & 4.4(2) & 0.30(1)\\ 
        &         &        & 18.2(4)  & 7.3(4) & 0.18(2)\\ 
        &         &        & 20.0(6)  & 8.8(5) & 0.14(3)\\ 
        &         &        & 23.0(8)  & 9.6(8) & 0.16(3)\\ \hline
$^{28}$Si& 21(6)  & 2.1(3) & 14.6(3)  & 6.0(3) & 0.44(4)\\ 
        &         &        & 24.6(8)  & 10.0(7) & 0.62(3)\\ 
\hline
\end{tabular}
%\end{ruledtabular}
\end{center}
\end{table}
The best fit GDR parameters for both the nuclei are given in Table~\ref{table1}.  It should be mentioned that the GDR lineshape is not expected to be very sensitive to the level density parameter~\cite{drc2010}. In the present case, the extracted GDR parameters corresponding to  $\it \tilde{a}$ = $\it A/$7 and $\it A/$8, are same within fitting errors. The effect of direct reactions like pre-equilibrium emission, incomplete fusion etc. is not considered, since it has been shown to be negligible  for $^{16}$O+$^{12}$C at the present beam energy~\cite{kundu2}.

Most noteworthy is the striking difference between the GDR spectra in two reactions leading to $^{31}$P and $^{28}$Si nuclei. Both $^{31}$P and $^{28}$Si, populated at the same initial excitation energy ($E^*\sim70$ MeV) and with angular momentum $<$$J$$>$ $\sim$ 21 $\hbar$ ($\pm$ 6 $\hbar$). The $J_C$ from systematics in Ref.~\cite{myers} are 19 $\hbar$ and 17 $\hbar$ for $^{31}$P and $^{28}$Si, respectively. 
While $^{31}$P spectrum shows the expected multicomponent character arising due to the Corioli's splitting with a distinct low energy peak at $\sim$ 9~MeV, the $\gamma$-ray spectrum of $^{28}$Si does not show evidence of the Jacobi shape transition. It can be seen from the Fig.~\ref{fig1} that $^{28}$Si  yield shows significant enhancement at higher folds as compared to $^{31}$P. Therefore, the measured fold distribution together with the fact that $J_C$ is lower for $^{28}$Si than that for $^{31}$P, makes the non-occurrence of Jacobi shape transition in $^{28}$Si very fascinating.
Further, the $\gamma$-ray spectrum of $^{28}$Si for $F\ge6$ corresponding to  $\langle J \rangle$ = 24 $\hbar$ ($\pm$ 6 $\hbar$) was also found to have same shape and no peak was visible around 10 MeV [see inset of Fig. 2(d)].

\begin{figure}
\includegraphics[width=.49\columnwidth, clip=true]{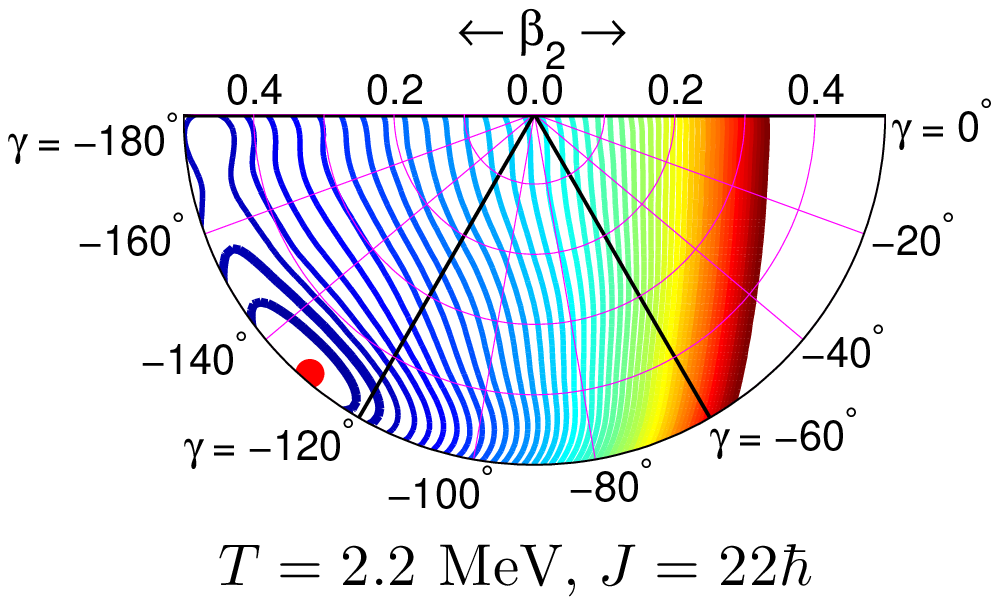}
\includegraphics[width=.49\columnwidth, clip=true]{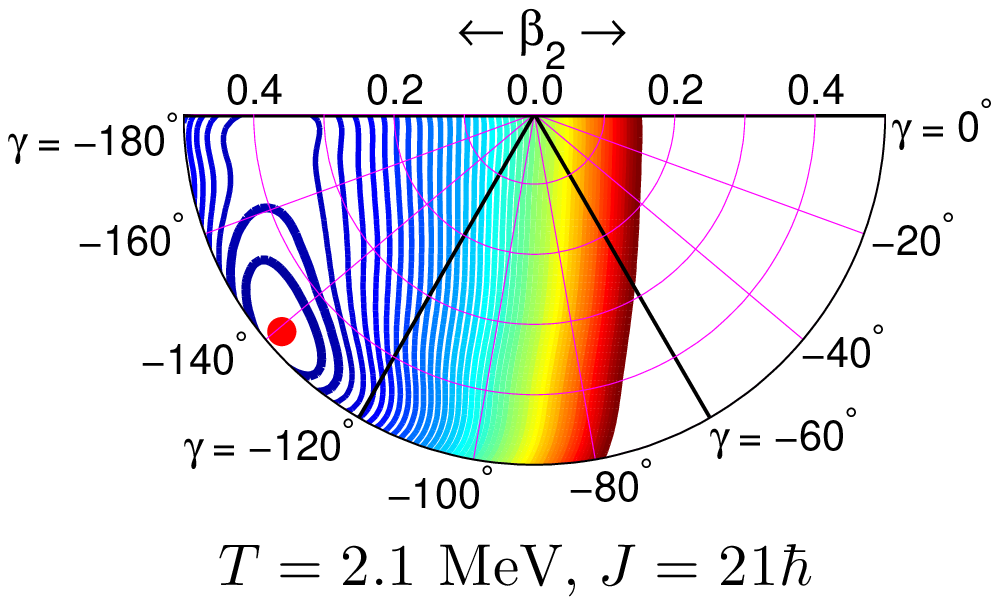}
\vskip-26pt
\hbox{\hspace{5pt}(a) \hspace{220pt}(b)}
\vskip10pt
\caption{\label{FES} (Color online) The free energy surfaces of (a) $^{31}$P and (b) $^{28}$Si for the measured $T$ and $J$ (contour line spacing is 0.2 MeV). Here, $\gamma=0^\circ$ ($-120^\circ$) represent the non-collective (collective) prolate shape and  $\gamma=-180^\circ$ ($-60^\circ$) represent the non-collective (collective) oblate shape.  The most probable shape is represented by a filled circle.}
\end{figure}

\begin{figure}
\begin{center}
\includegraphics[height=5.0 cm, width=7.6 cm]{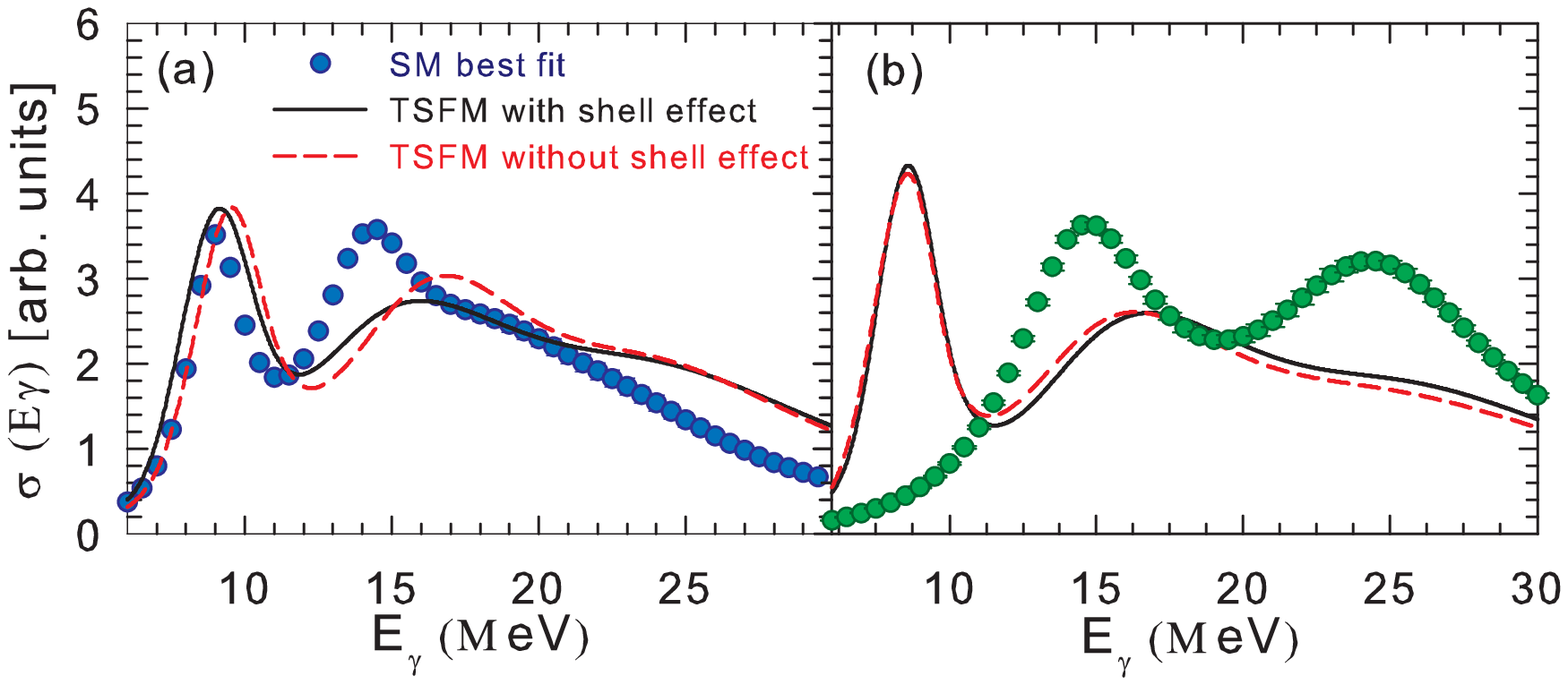}
\caption{\label{fig4} (Color online) The best fit  statistical model input cross-section (filled symbols) compared to TSFM calculations with (continuous line)  and without (dashed line) Shell effect.}
\end{center}
\end{figure}

The measured GDR strength functions are compared with thermal shape fluctuation model (TSFM) calculations corresponding to the measured $\langle T 
\rangle$ and  $\langle J \rangle$ values given in Table~\ref{table1}. The details of the TSFM calculation are discussed in the Refs. \cite{aru1,rhine1,rhine2,rhine3}, where 
shape fluctuations are treated by evaluating the expectation values  of the observables (over the deformation degrees of freedom) with their probability given by the Boltzmann factor [$\exp(-F/T)$].  The free energy ($F$) is calculated within a microscopic-macroscopic approach by tuning the angular frequency to get the desired $J$.
The calculated free energy surfaces (FES) are shown in Fig.~\ref{FES}, where it can be seen that the predicted equilibrium shapes for  both $^{31}$P and $^{28}$Si are similar and both the nuclei are  therefore expected to show similar behaviour -namely, the Jacobi shape transition. 
The calculated GDR cross-sections ($\sigma_{TSFM}$) are compared with the corresponding best fit statistical model calculation ($\sigma_{stat}$) in Fig.~\ref{fig4}. Since the absolute cross-section is not measured in the present experiment, $\sigma_{TSFM}$ was normalized to the total $\sigma_{stat}$ in the energy region of $E_{\gamma}$ = 7 - 25 MeV ~\cite{cghosh2016}. The variance in $\sigma_{stat}$ is calculated from the errors of the best fit parameters.  
The $^{31}$P data is in qualitative agreement with the TSFM predictions, but this is not the case for $^{28}$Si. 
The TSFM predicts a low energy component ($\sim$ 10 MeV) for $^{28}$Si, which is not corroborated by the data.
Further, the TSFM calculations  carried out without the shell effects are also shown in the same figure. It is seen that shell effects do not significantly affect the GDR cross-section at the measured $T$, $J$. The TSFM calculation does not include the pairing effect, which is expected to be negligible at $T\sim$ 2~MeV.
Therefore, it is evident that  the  GDR spectrum of $^{28}$Si at high $J$ is anamolous as compared to $^{31}$P and this discrepancy can not be understood in terms of TSFM or microscopic factors like shell or pairing effects. It should be noted that $^{16}$O+$^{12}$C reaction has entrance channel isopsin $T = 0$, which is expected to suppress the GDR yield ~\cite{behr} but is not expected to affect the shape of the GDR strength function.

Recently, similar observation- namely, the absence of Jacobi shape transition, was reported in $^{32}$S populated via $^{20}$Ne+$^{12}$C reaction~\cite{dipu2}. In both these cases ($^{32}$S and $^{28}$Si), projectile-target involve self-conjugate $\alpha$-cluster nuclei. As mentioned earlier, the reactions involving these nuclei are shown to exhibit orbiting phenomenon leading to formation of quasi-molecular states~\cite{tani,ichi}. The orbiting behaviour in $^{16}$O + $^{12}$C at $E_{lab}$ = 125 MeV was reported earlier in charged particle studies~\cite{kundu}.  In such molecular states, the configuration will have a two body rotor with mass concentrated on the periphery as opposed to a deformed nucleus with most of the mass at the center.  Hence, the moment of inertia corresponding to a  molecular resonance state is expected to be larger and consequently the angular frequency would be smaller. Thus, the formation of the dinuclear complex due to orbiting may suppress the Jacobi shape transition.
Further, in case of quasi-molecular resonances  there would be an  interplay of rotational motion of the dinuclear complex and vibrational motion of constituent nuclei, which would result in the fragmented strength ~\cite{cluster_gdr}. It should be pointed out that the net excitation energy as well as  effective  $T$ and $J$ for such a state can not be estimated in a simple manner. Moreover, the statistical model analysis or TSFM, which assumes a formation of equilibrated CN, is not suitable  to describe the data.   Detailed theoretical calculations are required to understand  whether the fragile correlations leading to molecular configurations  survive thermal fluctuations.  In addition, the role of possible binary shapes  on GDR, at large excitations needs to be investigated. 
It is important to note that the $^{32}$S  GDR data ~\cite{dipu1,dipu2} has been analyzed only within the statistical model framework.
However,  deformations deduced for both $^{32}$S, $^{28}$Si from the conventional statistical model analysis are  large  ($\beta_2 >0.6$) and  point towards the elongated structure.  It should be mentioned that the earlier data reporting the signature of orbiting in Ref.~\cite{kundu2} from the charged particle spectra, have shown the co-existence of molecular resonance states with equilibrated compound nucleus formation. In such a scenario, it is possible that high $J$ components of the entrance channel are predominantly contributing to the orbiting state and consequently the CN is formed with $J<J_C$. In the present experiment, $\gamma$-rays from CN with  $\langle J \rangle$ $\leq$ 15 $\hbar$ (corresponding to $F<$ 3) could not be unambiguously extracted. 

In summary,  the measurement of high energy $\gamma$-rays from the decay of giant dipole resonance in $^{31}$P nucleus and a self-conjugate $\alpha$-cluster nucleus $^{28}$Si, populated at same initial excitation energy and $\langle J \rangle$ $>$ J$_C$  was carried out to study the Jacobi shape transition. The measured GDR spectrum in  the decay of $^{31}$P  shows a distinct low energy component around 10 MeV, which is a clear signature of the Corioli's splitting in a highly deformed rotating nucleus. This first observation of the Jacobi shape transition in $^{31}$P, together with earlier results  in $A\sim40-50$ nulclei, show that Jacobi shape transition is a general feature of  nuclei in light mass region.
The observed GDR strength function in $^{31}$P  can be qualitatively explained by the TSFM. 
An anomalous behaviour is observed in the case of $^{28}$Si, where the GDR lineshape can be explained as 2-components to a prolate deformed nucleus, and does not exhibit signature of Jacobi shape transition.
Based on this data and similar recent results in $^{32}$S, it is proposed that the nuclear orbiting phenomenon exhibited by $\alpha$-cluster  nuclei, hinders the Jacobi shape transition.
%This data together with similar recent results in $^{32}$S are indicative of the orbiting phenomenon exhibited by $\alpha$-cluster.  It is proposed that the nuclear orbiting phenomenon exhibited by $\alpha$-cluster  nuclei, hinders the Jacobi shape transition.
The study of the GDR in self-conjugate $\alpha$-cluster CN populated through different entrance channels comprising $\alpha$-cluster and non-$\alpha$ cluster, would be important to understand the role of orbiting in nuclear structure. 
 The present experimental results suggest a possibility to investigate the nuclear orbiting phenomenon using high energy $\gamma$-rays as a probe. 
 
\section*{Acknowledgement}
We would like to thank Mr. M.S. Pose, Mr. K.S.~Divekar, Mr. M.E. Sawant, Mr. Abdul Quadir, Mr. R. Kujur for help with experimental setup, Mr. R.D. Turbhekar for target preparation and the PLF staff for the smooth operation of the accelerator. AKRK acknowledges the financial support from the DST-INSPIRE Faculty program (India) and RIKEN Supercomputer HOKUSAI GreatWave System for the numerical calculations. PA acknowledges financial support from the SERB (India), DST/INT/POL/P-09/2014.


\begin{thebibliography}{99}

\bibitem{beri} R Beringer and W J Knox, Phys. Rev.  121 (1961) 1195.
\bibitem {myers} W. D. Meyers and W. J. Swiatecki,  Acta. Phys. Pol. B  32 (2001) 1033.
\bibitem {dudek}Mazurek  et al., Phys. Rev. C  91 (2015) 034301.
\bibitem{kici} M. Kicihska-Habior  et al., Phys. Lett. B.  308 (1993) 225.
\bibitem{maj} A. Maj  et al., Nucl. Phys. A  731 (2004) 319.
\bibitem{drc1} D. R. Chakrabarty  et al., Phys. Rev. C  85 (2012) 044619.
\bibitem{dipu1} Deepak Pandit  et al., Phys. Rev. C  81 (2010) 061302R.
\bibitem{DRCrev} D. R. Chakrabarty, N. Dinh Dang and V. M. datar, Eur. Phys. J. A  52 (2016) 143. 
\bibitem{ward} D. Ward  et al., Phys. Rev. C  66 (2002) 024317.
\bibitem{freer1} M. Freer, Rep. Prog. Phys.  70 (2007) 2149.
\bibitem{von} W. von Oertzen, M. Freer, and Y. Kanada-En’yo, Phys. Rep.  432 (2006) 43.
\bibitem{Aradhana} A. Shrivastava  et al., Phys. Lett. B.  718 (2013) 931.
\bibitem{ebran} J. P. Ebran, E. Khan, T. Niksic and D. Vretenar, Nature  487 (2012) 341.
\bibitem{hoyle} F Hoyle, Astrophys. J. Suppl. Ser.  1 (1954) 121.
\bibitem{rana} T. K. Rana  et al., Phys. Rev. C  88 (2013) 021601R.
\bibitem{freer2} Martin Freer and H.O.U Fyndo, Prog. Part. Nucl. Phys.  78 (2014) 2149.
\bibitem{sand} S. J. Sanders, A. Szanto de Toledo, and C. Beck, Phys. Rep.  311 (1999) 487.
\bibitem{kundu} S.  Kundu  et al., Phys. Rev. C  78 (2008) 044601.
\bibitem{tani} Yasutaka Taniguchi  et al., Phys. Rev. C  80 (2009) 044316.
\bibitem{ichi} T. Ichikawa, Y. Kanada-Enyo, and P. Moller, Phys. Rev. C  83 (2011) 054319.
\bibitem{beck} C. Beck  et al., AIP conference proceedings  1098 (2009) 207.
\bibitem{salsac} M.D. Salsac  et al.,Nucl.  Phys. A  801 (2008) 1.
\bibitem{dipu2} Deepak Pandit  et al., Phys. Rev. C  95 (2017) 034301.
\bibitem{lamps} http:/www.tifr.res.in/$\sim$pell/lamps.html
\bibitem{ghosh17} C. Ghosh  et al.,  Phys. Rev. C  96 (2017) 014309.
\bibitem{chakraborty2} D. R. Chakrabarty  et al., Nucl. Instrum. Methods Phys. Res., Sect. A  560 (2006) 546.
\bibitem{perey1} C. M. Perey and F. G. Perey, At. Data Nucl. Data Tables  17 (1976) 1.
\bibitem{perey2} F. G. Perey, Phys. Rev.  131 (1963) 745.
\bibitem{mcfa} L. Mcfadden and G. R. Satchler, Nucl. Phys. A  84 (1966) 177.
\bibitem{igna} A. V. Ignatyuk, G. N. Smirenkin, and A. S. Tishin, Sov. J. Nucl.Phys.  21 (1975) 255 [Yad. Fiz.  21 (1975) 485].
\bibitem{debasish} Debasish Mondal  et al., Phys. Lett. B.  763 (2016) 422.
\bibitem{chakraborty3} D. R. Chakrabarty  et al., Nucl. Phys. A  770 (2006) 126.
\bibitem{nifnecker} H. Nifennecker and J. A. Pinston, Annu. Rev. Nucl. Part. Sci.  40 (1990) 113.
\bibitem{cori-fit} K. Neegard  et al., Phys. Lett. B.  110 (1982) 7.
\bibitem{drc2010} D. R. Chakrabarty  et al., J. Phys. G: Nucl. Part. Phys.  37 (2010) 055105.
\bibitem{kundu2} S.  Kundu  et al., Phys. Rev. C  87 (2013) 024602.
\bibitem{aru1} P. Arumugam, G. Shanmugam and S. K. Patra., Phys. Rev. C  69 (2004) 054313.
\bibitem{rhine1} A. K. Rhine Kumar, P. Arumugam and N. Dinh Dang., Phys. Rev. C  90 (2014) 044308.
\bibitem{rhine2} A. K. Rhine Kumar, P. Arumugam and N. Dinh Dang., Phys. Rev. C  91 (2015) 044305.
\bibitem{rhine3} A. K. Rhine Kumar and P. Arumugam., Phys. Rev. C  92 (2015) 044314.
\bibitem{cghosh2016} C. Ghosh  et al.,  Phys. Rev. C  94 (2016) 014318.
\bibitem{behr} J. A. Behr  et al., Phys. Rev. Lett.  70 (1993) 3201.
\bibitem{cluster_gdr} W. B. He, Y. G. Ma, X. G. Cao, X. Z. Cai and G. Q. Zhang, Phys. Rev. Lett.  112 (2014) 032506.


\end{thebibliography}
\end{document}